\documentstyle[12pt,epsfig]{article}
\def \app{D_{\pi \pi}}
\def \dkk{D_{KK}}

\def \bea{\begin{eqnarray}}
\def \beq{\begin{equation}}
\def \bo{B^0}
\def \ko{K^0}
\def \bs{B_s}
\def \obs{\overline{B}_s}

\def \cn{Collaboration}
\def \cpp{C_{\pi \pi}}
\def \ckk{C_{KK}}
\def \eea{\end{eqnarray}}
\def \eeq{\end{equation}}
\def \ite{{\it et al.}}

\def \lkk{\lambda_{KK}}
\def \ob{\overline{B}^0}
\def \ok{\overline{K}^0}
\def \rpp{R_{\pi \pi}}
\def \rkk{R_{KK}}

\def \spp{S_{\pi \pi}}
\def \skk{S_{KK}}
\begin{document}

\begin{flushright}
TECHNION-PH-2002-14\\
EFI 02-68 \\
hep-ph/0203158 \\
March 2002 \\
\end{flushright}

\renewcommand{\thesection}{\Roman{section}}
\renewcommand{\thetable}{\Roman{table}}
\centerline{\bf WEAK PHASE $\gamma$ FROM $\bs(t) \to K^+K^-$}
\medskip
\centerline{Michael Gronau}
\centerline{\it Physics Department, Technion -- Israel Institute of Technology}
\centerline{\it 32000 Haifa, Israel}
\medskip
\centerline{Jonathan L. Rosner}
\centerline{\it Enrico Fermi Institute and Department of Physics}
\centerline{\it University of Chicago, Chicago, Illinois 60637}
\bigskip

\begin{quote}

We re-examine the time-dependent rates of $\bs(t) \to K^+ K^-$ and 
$\obs(t) \to K^+ K^-$, including a lifetime difference between neutral $\bs$ 
mass eigenstates. The two rates, normalized by the rate of $\bs \to \ko \ok$, 
are used to obtain ambiguity-free information on a strong phase and on the 
weak phase $\gamma$. We discuss the sensitivity of extracting $\gamma$ to 
the measured quantities, and find that an error of $\pm 10^\circ$ in $\gamma$
is possible for a sample of several thousand $\bs(t) \to K^+ K^-$
decays.  This study is complementary to a recent similar analysis of the U-spin
related decays $\bo(t) \to \pi^+\pi^-$ and $\ob(t) \to \pi^+\pi^-$.

\end{quote}

\leftline{\qquad PACS codes:  12.15.Hh, 12.15.Ji, 13.25.Hw, 14.40.Nd}

\section{Introduction}

Time-dependent CP asymmetry measurements in neutral $B$ meson decays 
provide useful tests for the standard model, which attributes  CP 
violation to a single phase in the Cabibbo-Kobayashi-Maskawa (CKM) matrix.
The potential variety of $\bo$ and $\bs$ decay modes permits one to 
overconstrain the CKM unitarity triangle involving the three angles 
$\alpha = \phi_2,~\beta \equiv \phi_1$ and $\gamma = \phi_3$ \cite{PDG}.
The observation of CP violation in $B^0(t) \to J/\psi K_S$ \cite{betaBa,betaBe},
interpreted in terms of $\sin 2\beta$, passed this test successfully. 
First results of a CP asymmetry measurement in $\bo(t) \to \pi^+\pi^-$,  
consistent with a zero asymmetry, though involving large errors, were reported 
in \cite{CSBa}.
Updates, implying world average asymmetries which are nonzero at about a
$2.3\sigma$ level, were presented very recently \cite{Moriond1,Moriond2}. 
The two measured quantities, coefficients of terms involving 
$\sin \Delta m t$ and $\cos \Delta m t$ in the time-dependent asymmetry, were 
studied recently \cite{GR} as functions of relevant strong and weak phases. 
It was shown how discrete ambiguities in a strong phase can be resolved in 
order to learn the weak phase $\alpha$. 

Approximate flavor SU(3) symmetry of strong interactions can be used to
relate a variety of $B$ and $\bs$ decays to two charmless pseudoscalar
mesons \cite{GHLR}. Thus, the process $\bo \to\pi^+\pi^-$ is related to 
$\bs \to K^+K^-$ in the limit of U-spin symmetry involving an interchange of 
$d$ and $s$ quarks. This symmetry implies equal CP rate differences in any 
pair of U-spin related processes \cite{Uspin}, and in particular in this
pair of decay modes. The U-spin relation between
$\bo \to\pi^+\pi^-$ and $\bs \to K^+K^-$ was used in \cite{FL} to suggest 
a simultaneous determination of the two weak phases $\beta$ and $\gamma$ by 
combining information from the asymmetries in these two processes.
By measuring for both channels a $\cos \Delta mt$ and a 
$\sin\Delta mt$ term one has four equations for four unknowns: $\beta$ (which 
can also be considered as given from direct measurements), $\gamma$, the ratio 
of penguin to tree amplitudes in $\bo\to \pi^+\pi^-$, and the relative strong 
phase between these two amplitudes. It was noted \cite{FL} that factorized 
U-spin breaking effects cancel in the double ratio of penguin to tree
amplitudes and in the difference between the two relative strong phases
occuring in the two processes.  This reduces the intrinsic theoretical
uncertainty of this method.

Several earlier suggestions for measuring weak phases in $B_s \to K^+ K^-$ 
were made in \cite{KLY,FL2}, by combining this decay within flavor SU(3) 
with information from other processes, involving $\bs \to \ko\ok,~\bs 
\to K^-\pi^+, ~\bo \to K^+\pi^-$ and $B^+ \to \ko \pi^+$.  Usually, these
methods require some assumption about SU(3) breaking. Also, as a result of
the large number of required measurements, which are expressed in terms 
of trigonometric functions of strong and weak phases, one
encounters several discrete ambiguities in the solutions for weak phases. 

In the present paper we wish to re-examine more carefully the simplest
among all these suggestions \cite{KLY}, which does not require full SU(3)
but only isospin symmetry. The method is based primarily on the time-dependent 
rate of $\bs(t) \to K^+K^-$ and its charge-conjugate, normalized by the 
time-integrated rate of $\bs \to \ko\ok$.  We will also include effects due to
a lifetime difference between neutral $\bs$ mass eigenstates which were 
neglected in \cite{KLY}. 

A major purpose of this study will be to show how to 
resolve certain discrete ambiguities in strong and weak phases in order to 
determine unambiguously the weak phase $\gamma$. One of the outputs of the 
above measurements is the ratio of tree to penguin amplitudes in $\bs 
\to K^+K^-$. This ratio is related by U-spin to the corresponding ratio in 
$\bo \to \pi^+\pi^-$, for which some information already exists. We will
use this information to demonstrate the sensitivity of this method.

We introduce notations for $\bs \to K^+K^-$ and $\bs \to \ko\ok$ in Sec.\ II,
where we explain some necessary assumptions and ways of testing these 
assumptions. The general time-dependence of $\Gamma(\bs(t) \to K^+K^-)$, 
normalized by $\Gamma(\bs \to \ko\ok)$, involves four measurable quantities 
$\rkk,~\ckk,~\skk$ and $\dkk$, for which expressions are derived in 
Sec.\ III in terms of a ratio of tree to penguin amplitude, a strong phase 
difference and the weak phase $\gamma$. We discuss the resolution of
ambiguities associated with the strong phase.  Solutions for $\gamma$ in terms
of the measurables are studied in Sec.\ IV
in order to evaluate the sensitivity of the method. We conclude in Sec.~V.

\section{Notation and assumptions}

In Ref.\ \cite{GR}, studying $\bo \to \pi^+\pi^-$ in terms of tree and 
penguin amplitudes,
\beq\label{pipi}
A(\bo \to \pi^+ \pi^-) = -(|T|e^{i \delta_T} e^{i \gamma} +
|P| e^{i \delta_P})~~~,
\eeq
we used the convention in which top quark contributions are integrated out 
in the short-distance effective Hamiltonian, and the unitarity relation 
$V^*_{ub} V_{ud} + V^*_{cb} V_{cd} = - V^*_{tb} V_{td}$ is used. The 
$V^*_{ub} V_{ud}$ piece of the penguin operator is included in the tree 
amplitude. We adopt the same convention for $\bs \to K^+ K^-$, such that
\beq\label{bs}
A(\bs \to K^+ K^-) = -(|T'|e^{i \delta_T'} e^{i \gamma} -
|P'| e^{i \delta_P'})~~~.
\eeq
In the U-spin symmetry limit one has \cite{Uspin, FL}
\beq\label{U}
\frac{|T'|}{|T|} = \frac{V^*_{ub}V_{us}}{V^*_{ub}V_{ud}} = \frac{\lambda}
{1 - \lambda^2/2}~~,~~~~\frac{|P'|}{|P|} = 
-\frac{V^*_{cb}V_{cs}}{V^*_{cb}V_{cd}} = \frac{1 - \lambda^2/2}{\lambda}~~,
~~~~\delta_T' = \delta_T~~,~~~~\delta_P' = \delta_P~~~,
\eeq
where $\lambda = 0.22$ \cite{PDG}. 

We note in passing that Eq.~(\ref{bs}) by itself is a general expression of 
the decay amplitude in the CKM framework, and involves no assumption.
The amplitude for $\obs \to K^+ K^-$ is simply 
obtained by changing the sign of the weak phase $\gamma$.

The amplitude for $\bs \to \ko \ok$ will be assumed to have the simple form
\beq\label{bso}
A(\bs \to \ko \ok) = -|P'| e^{i \delta_P'}~~~.
\eeq
The QCD penguin terms in $\bs$ decays to charged and neutral kaons are equal 
by isospin. However, the form (\ref{bso})  which assumes equal overall penguin 
contributions in the two processes and no tree amplitude (with weak phase 
$\gamma$) in $\bs \to \ko\ok$, involves two approximations. First, we neglect a 
small color-suppressed electroweak penguin contribution in $\bs \to K^+K^-$
\cite{GHLR}. Using factorization, such contributions are estimated to be at
a level of a few percent \cite{FM}. Tests for the magnitude of the electroweak 
penguin contribution in $\bs \to K^+ K^-$ were proposed in \cite{FL2}. In 
Eq.~(\ref{bso}) one also disregards the possibility of enhanced rescattering 
effects in $\bs \to \ko \ok$, typically of order 
$|V^*_{ub}V_{us}/V^*_{cb}V_{cs}| \simeq 2\%$ without such
enhancement. Such enhanced effects could, in principle, induce a sizable term 
with phase $\gamma$. One test for such a term would be observing a CP asymmetry 
between this process and its change conjugate. Other indirect tests involve
rate measurements of hadronic decays such as $\bs \to \pi^+\pi^-$ and $\bo 
\to K^+K^-$, which are very rare in the absence of rescattering \cite{BGR}. 

Since in the above approximation $\bs \to \ko \ok$ involves a single weak
phase, we expect no CP asymmetry in this decay.
We denote by $\rkk$ the ratio of charge averaged rates for $\bs~({\rm or}~\obs)
\to K^+K^-$ and $\bs~({\rm or}~\obs) \to \ko \ok$,
\beq\label{rkk}
\rkk \equiv \frac{{\cal B}(\bs \to K^+ K^-) + {\cal B}(\obs  \to K^+ K^-)}
{{\cal B}(\bs \to \ko \ok) + {\cal B}(\obs  \to \ko \ok)}~~~.
\eeq

The time-dependent decay rate of $\bs|_{\rm initial} \to K^+ K^-$ is given by
\cite{isi}
\bea\label{Gamma(t)}
\Gamma(\bs(t) \to K^+K^-) \propto e^{-\Gamma_s t} \times
\left [ \cosh (\Delta\Gamma_s t/2) - \dkk\sinh (\Delta\Gamma_s t/2)
\right. \nonumber\\
\left.~+ \ckk\cos(\Delta m_s t) - 
\skk\sin(\Delta m_s t) \right ]~~~,
\eea
where $\Gamma_s \equiv (\Gamma_H+\Gamma_L)/2$,  and $\Delta m_s\equiv m_H - 
m_L$ and $\Delta\Gamma_s \equiv \Gamma_L - \Gamma_H$ are the mass and width 
differences between the two $\bs$ mass eigenstates. In the standard model one 
expects $\Delta\Gamma_s > 0$, and estimates imply that 
$\Delta\Gamma_s/\Gamma_s$ can be as large as about 15$\%$ \cite{Lenz}.
An expression similar to (\ref{Gamma(t)}), in which the 
$\ckk$ and $\skk$ terms occur with opposite signs, describes the 
rate of $\obs|_{\rm initial} \to K^+ K^-$. 

The coefficients $\ckk,~\skk$ and $\dkk$, obeying
\beq
\ckk^2 + \skk^2 + \dkk^2 = 1~~~,
\eeq
are defined by
\beq \label{CSDkk}
\ckk \equiv \frac{1 - |\lkk|^2}{1 + |\lkk|^2}~~,~~~
\skk \equiv \frac{2 {\rm Im}(\lkk)}{1 + |\lkk| ^2}~~,~~~
\dkk \equiv \frac{2 {\rm Re}(\lkk)}{1 + |\lkk|^2}~~~,
\eeq
where
\beq
\lkk \equiv \frac{A(\obs \to K^+ K^-)}{A(\bs \to K^+ K^-)}~~~.
\eeq
A very small overall phase from $\bs-\obs$ mixing, 
${\rm arg}(V^*_{tb}V_{ts}/V^*_{cb}V_{cs})^2 \sim \lambda^2$, was neglected.

Defining $\delta' \equiv \delta_P' - \delta_T'$, where we use a convention 
in which $-\pi \le \delta' \le \pi$, one finds
\beq
\lkk = \frac{1 - |T'/P'| e^{-i (\delta' + \gamma)}}
{1 - |T'/P'| e^{-i (\delta' - \gamma)}}~~~.
\eeq
This expression simplifies to a pure phase in the two special cases when
$\delta'=0$ or $\pi$
\beq
\lkk = e^{2 i \zeta}~~,~~~~
\zeta = \left\{ \begin{array}{c} \arctan \frac{|T'/P'| \sin\gamma}
{1 - |T'/P'| \cos\gamma}~~~(\delta' = 0) \cr
- \arctan\frac{|T'/P'| \sin\gamma}{1 + |T'/P'| \cos \gamma}~~~(\delta' =
\pi)~~~. \end{array} \right.
\eeq
In such cases $\ckk = 0,~\skk = \sin 2\zeta,~\dkk = \cos 2\zeta$.

In the next section we will study the dependence of the measurables defined in 
Eqs.~(\ref{rkk}) and (\ref{CSDkk}) on $|T'/P'|,~\delta'$ and $\gamma$.

\section{Dependence of observables on $\delta'$ and $\gamma$}

Expressions for $\rkk,~\ckk,~\skk$ and $\dkk$ in terms of $|T'/P'|,~\delta'$
and $\gamma$ are readily derived:
\bea
\rkk &=& 1 - 2|T'/P'|\cos\delta'\cos\gamma + |T'/P'|^2~~~,\\
\ckk &=& -2|T'/P'|\sin\delta'\sin\gamma/\rkk~~~,\\
\skk &=& [2|T'/P'|\cos\delta'\sin\gamma - |T'/P'|^2\sin 2\gamma]/\rkk~~~,\\
\dkk &=& [1 - 2|T'/P'|\cos\delta'\cos\gamma + |T'/P'|^2\cos 2\gamma]/\rkk~~~.
\eea 

There are some interesting differences between the present case of $B_s
\to K^+ K^-$ and that of $B^0 \to \pi^+ \pi^-$ discussed in Ref.\ \cite{GR}.

\begin{enumerate}

\item Assuming ${\rm sign}(\sin\delta') = {\rm sign}(\sin\delta)$, the 
quantities $\ckk$ and $\cpp$ are opposite in sign. In the U-spin 
symmetry limit, the corresponding CP rate differences are equal in 
magnitude and have opposite signs.  

\item In contrast to the case of $\spp$,
which measures $\sin 2\alpha$ in the limit $|P/T| \to 0$,
there is no term independent of the ratio of the smaller amplitude to 
the larger amplitude; $\skk$ vanishes in the limit $|T'/P'| \to 0$.

\item Whereas $\app$ is expected to be large and negative, $\dkk$ is
predicted to be large and positive.  Its maximal value is +1.

\item While one expects $\rpp > 1$ for $\cos \gamma > 0$ and $\cos
\delta \simeq 1$, one expects $\rkk < 1$ for $\cos \delta' \simeq 1$.

\item All equations involve only the CKM phase $\gamma$, while for
$B^0 \to \pi^+ \pi^-$ the relations are not as simple because they
contain both $\alpha$ and $\gamma$
(or $\beta$ and $\gamma$.) This should not be considered a disadvantage,
since $\beta$ is known rather precisely from direct measurements.

\end{enumerate}

The first three expressions for $\rkk,~\ckk$ and $\skk$ agree with results 
obtained in \cite{KLY} using different notations.  One may regard these three
equations as specifying the three unknowns $|T'/P'|,~\delta'$ and $\gamma$.
The quadratic equations for $|T'/P'|$ and the trigonometric expressions in
terms of strong and weak phases yield solutions which involve several discrete
ambiguities. 

Whereas the magnitude of $\dkk$ is specified by $\ckk$ and $\skk$, which are 
likely to be more easily measured, its sign could in principle resolve some of 
these ambiguities. $\dkk$ may change sign as a function of $\delta'$ and
$\gamma$ if $|T'/P'|$ is sufficiently large.  For instance, in \cite{KLY} this
ratio was taken to be $8/9$, which would permit a change of sign in $\dkk$.
As we will show now, there exists already indirect information on $|T'/P'|$
from nonstrange $B$ decays which implies $|T'/P'| = 0.184 \pm 0.043$,
which is much smaller than assumed in \cite{KLY}. With such a small value of 
$|T'/P'|$, $\dkk$ must be positive and large
in the CKM framework. A negative value would signify new physics.   

Using Eq.\ (\ref{U}), we find to leading order in $\lambda^2$,
\beq\label{T'/P'}
\frac{|T'|}{|P'|} = \frac{\lambda^2}{1 - \lambda^2}\frac{|T|}{|P|}~~~.
\eeq
This relation, which is precise in the U-spin symmetry limit, may be affected 
by U-spin breaking. However, as noted in \cite{FL}, such effects are expected 
to be small in the ratio of ratios $|T'/P'|:|T/P|$ if one assumes approximate 
factorization. We will neglect nonfactorizable corrections in our evaluation of 
$|T'/P'|$. The ratio $|P/T|$ was estimated in Refs.\ \cite{GR01} and 
\cite{LR}, applying SU(3) with SU(3) breaking to experimental data on 
$B^+ \to K^0 \pi^+$ (a process dominated by the penguin amplitude), and
applying data on $B \to \pi l \nu$ related to the tree amplitude in 
$\bo \to \pi^+\pi^-$ in the factorization approximation.  We shall use the 
result of Ref.\ \cite{GR01},
$|P/T| = 0.276 \pm 0.064$.  Ref.\ \cite{BBNS}, based on  explicit
calculations in QCD-improved factorization, found a very similar value
$|P/T| = 0.285 \pm 0.076$. Thus, we find from Eq.~(\ref{T'/P'}) 
\beq\label{T'/P'number}
\frac{|T'|}{|P'|} = 0.184 \pm 0.043~~~.
\eeq
As mentioned, this value allows only positive values of $\dkk$,
which cannot provide new information for reducing discrete ambiguities.

Let us discuss ambiguities related to the strong phase $\delta'$, which
can be positive or negative, and can lie either in the range $0 < |\delta'| 
\le \pi/2$ or in $\pi/2 < |\delta'| \le \pi$. Calculations based on 
QCD-improved factorization \cite{BBNS} may indicate that this phase (related 
by U-spin to the phase $\delta \equiv \delta_P - \delta_T$ defined in
(\ref{pipi})) is small and has a definite sign; a value $\delta \simeq
10^\circ$ is found in Ref.\ \cite{BBNS}.  Other 
perturbative QCD calculations \cite{KLS} imply a larger negative phase 
$\delta$; correspondingly, Ref.\ \cite{Chen} finds $\delta' \simeq
-27^\circ$ when the phase is expressed in our convention.  It will be 
important to check these predictions.

In order to resolve ambiguities in $\delta'$, we note that $\rkk,~\skk$ and 
$\dkk$ are even in $\delta'$ while $\ckk$ is odd in $\delta'$.  First, consider
$\rkk$. Since current constraints on CKM parameters \cite{Beneke,Hocker,Ciu}
imply $\gamma <~\pi/2$ or $\cos\gamma > 0$, a value $\rkk < 1$ would imply
$\cos\delta' > 0$, namely $0 < |\delta'| < \pi/2$.  Furthermore, a value of
$\rkk$ below 1 permits one to set a significant bound on $\gamma$ which is
independent of $\delta'$ \cite{FM,GR},
\beq
\sin^2\gamma \le \rkk~~~.
\eeq  

Second, consider $\ckk$.  Within the CKM framework $\sin\gamma > 0$. Therefore, 
a nonzero measurement of $\ckk$ will specify the sign of $\delta'$. 
There exists an absolute bound on $\ckk$, $|\ckk| \le 2|T'/P'|/(1 + 
|T'/P'|^2)$, which becomes stronger for a {\it given} value of $\gamma$:
\beq
|\ckk| \le \frac{2|T'/P'| |\sin\gamma|}{\sqrt{(1 + |T'/P'|^2)^2
- 4|T'/P'|^2 \cos^2\gamma}}~~~.
\eeq

Examining $\skk$, we note that its dominant term is $2|T'/P'|\cos\delta'\sin
\gamma$. Hence, neglecting the term quadratic in $|T'/P'|$, $\skk \ge 0$ for $0
< |\delta'| \le \pi/2$, and $\skk < 0$ for $\pi/2 < |\delta'| \le \pi$, which
resolves the same ambiguity in $\delta'$ as in $\rkk$. In this approximation,  
$\rkk\skk$ is bound by $|\rkk\skk| \le 2|T'/P'||\sin\gamma|$ for a given 
value of $\gamma$.

\section{Solutions for $\gamma$}

Combining the equations for $\rkk$ and $\ckk$, one can eliminate the strong
phase $\delta'$
\beq\label{rkk1}
\rkk = 1 + |T'/P'|^2 \pm \sqrt{4|T'/P'|^2\cos^2\gamma - 
(\rkk\ckk)^2\cot^2\gamma}~~~.
\eeq
This equation can be inverted to obtain a quadratic equation for $\sin^2\gamma$
in terms of $\rkk$ and $\ckk$ \cite{similar}:
\bea
& &4|T'/P'|\sin\gamma = \nonumber \\
& &\pm \{[(1 + |T'/P'|)^2 - \rkk(1 + \ckk)][\rkk(1 - \ckk) -
(1 -|T'/P'|)^2] \}^{1/2} \nonumber\\
& &\pm \{[(1 + |T'/P'|)^2 - \rkk(1 - \ckk)][\rkk(1 + \ckk) - 
(1 - |T'/P'|)^2] \}^{1/2}.
\eea
As mentioned, only positive solutions of $\sin\gamma$ apply within the CKM 
framework.

While the value of $|T'/P'|$ in Eq.~(\ref{T'/P'number}) was obtained indirectly 
from nonstrange $B$ decays, a direct cross check in $\bs \to K \bar K$ can 
be achieved by using information from $\skk$. One obtains
\beq\label{rkk2}
\rkk(1 + \skk\cot\gamma) = 1- |T'/P'|^2\cos 2\gamma~~~.
\eeq
Alternatively, neglecting the term in $\skk$ quadratic in $|T'/P'|$, one 
gets a useful approximation,
\beq
4|T'/P'|^2 \sin^2\gamma \approx \rkk^2(\ckk^2 + \skk^2)~~~.
\eeq

\begin{figure}[t]
\centerline{\epsfysize = 4.9 in \epsffile{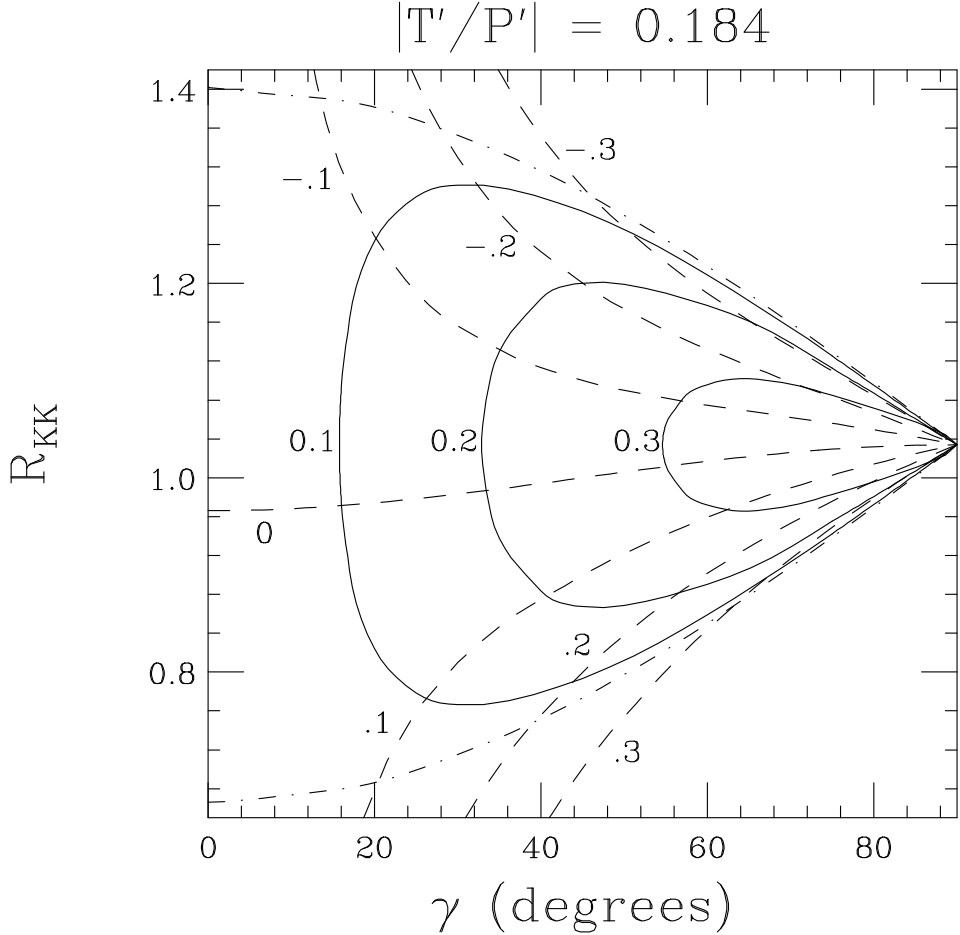}}
\caption{Dependence of $\rkk$ on $\gamma$ for fixed values of $|\rkk\ckk|$
(labels on solid curves).  Dot-dashed curves correspond to 
$\ckk = 0$.  Dashed curves (labeled by values of $\rkk \skk$)
show values of $\rkk$ for fixed values of $\rkk \skk$.  Only the
range $0 \le \gamma \le 90^{\circ}$ is shown since all equations are
invariant under $\gamma \to \pi - \gamma$, $\skk \to - \skk$.  These curves
were plotted for $|T'/P'| = 0.184$.}
\end{figure}

Graphical solutions of Eqs.~(\ref{rkk1}) and (\ref{rkk2}) can be used to
demonstrate the sensitivity of the extracted value of $\gamma$ to 
the measured quantities. In Fig.~1 we use these two equations
to plot $\rkk$ as a function of $\gamma$ for several values of $\rkk\ckk$ and 
for several values of $\rkk\skk$. For $|T'/P'|$ we adopt the
central value in Eq.~(\ref{T'/P'number}), $|T'/P'| = 0.184$.
Figs.~2 and 3 show the effects of varying $|T'/P'|$ by $\pm 1 \sigma$.

The curves in Figs.\ 1--3 indicate that $\gamma$ can be specified by
measurements of $|T'/P'|$, $\rkk$, $\skk$, and $\ckk$.  The equations are
overconstrained, allowing the elimination of some discrete ambiguities.
Usually, measurements of $\rkk$ and $\ckk$ are seen to leave a twofold
ambiguity in the solution for $\gamma$ within the physical CKM range, $0 <
\gamma < \pi/2$, and another twofold ambiguity outside this range.  
This fourfold ambiguity is resolved by $\skk$. 

\begin{figure}
\centerline{\epsfysize = 4.9 in \epsffile{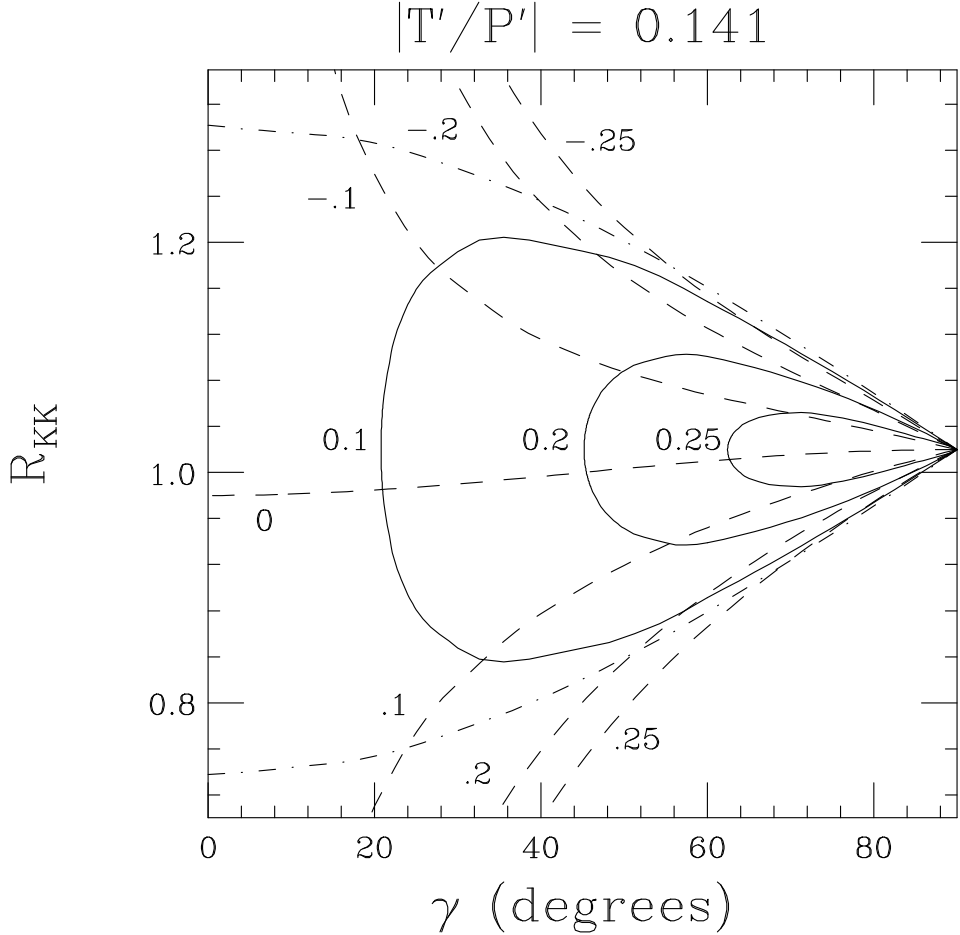}}
\caption{Dependence of $\rkk$ on $\gamma$ for fixed values of $|\rkk\ckk|$
(solid curves) and $\rkk \skk$ (dashed curves).  Same as Fig.\ 1 except
$|T'/P'| = 0.141$.}
\end{figure}

As an example, assume that $|T'/P'| = 0.184$, and
let us imagine that we observe $\rkk = 0.8$ and $|\rkk \ckk|
= 0.1$.  Acceptable solutions are $\gamma \simeq \gamma_1 \equiv 22^\circ$,
$\gamma_2 \equiv 47^\circ$, $\pi - \gamma_1$, and $\pi - \gamma_2$.  These
values correspond, respectively, to $\rkk \skk \simeq 0.07,~0.22,-0.07,-0.22$.
Thus, even given perfect measurements of (say) $\rkk$ and $\ckk$, one will
need to determine $\rkk \skk$ to an accuracy of $\pm 0.05$ or better in
order either to resolve the discrete ambiguity or to encounter an
inconsistency. 
For $|T'/P'| = 0.141$ these measurements of $\rkk$ and $\ckk$ lead to no 
solution of $\gamma$, and for $|T'/P'| = 0.227$ the ambiguity in $\gamma$ 
can be resolved with a less accurate measurement of $\skk$. 
This indicates that at least several hundred flavor tagged $B_s \to K^+ K^-$
decays would be needed for the present method to be useful. 
This estimate ignores background and resolution factors. 
A less optimistic estimate is based on extrapolation of the BaBar
result for $B^0 \to \pi^+ \pi^-$ \cite{CSBa}, in which a sample of 65
flavor tagged
events led to errors of about $\pm (0.5,0.6)$ in $(\cpp,\spp)$. If these
errors scale as the inverse square root of the number of events, it may be
necessary to obtain several thousand tagged 
$B_s \to K^+ K^-$ decays to achieve errors of $\pm 0.05$ in $\ckk$ and $\skk$. 

\begin{figure}
\centerline{\epsfysize = 4.9 in \epsffile{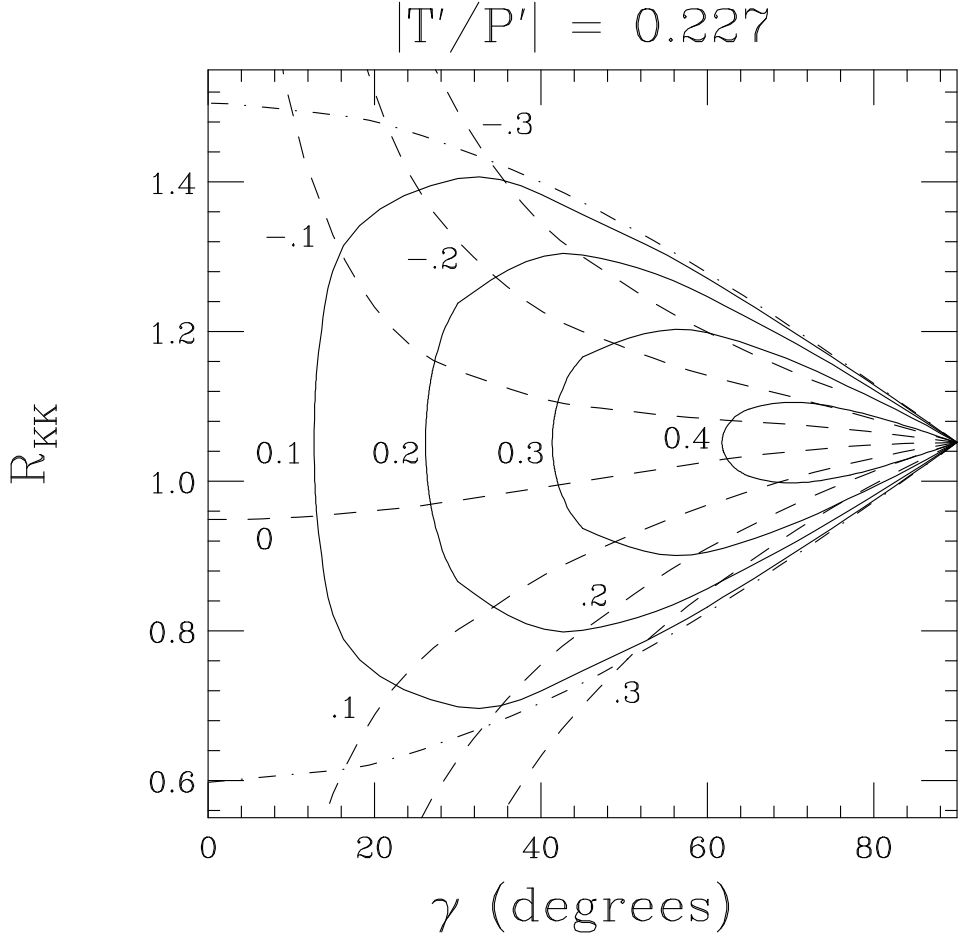}}
\caption{Dependence of $\rkk$ on $\gamma$ for fixed values of $|\rkk\ckk|$
(solid curves) and $\rkk \skk$ (dashed curves).  Same as Fig.\ 1 except
$|T'/P'| = 0.227$.}
\end{figure}

Equations (\ref{rkk1}) and (\ref{rkk2}) can be used to obtain both $|T'/P'|$ 
and $\gamma$. Eliminating $|T'/P'|$, 
{\em one expresses $\gamma$ in terms of measurable quantities},
\bea\label{RS}
\rkk\skk & = & -\frac{1}{2}\sin 4\gamma - (\rkk - 1)\sin 2\gamma 
\nonumber\\
& \pm & \cos 2\gamma \sqrt{\sin^2 2\gamma + 4(\rkk - 1)\sin^2\gamma - 
(\rkk\ckk)^2}~~~,
\eea
while $|T'/P'|$ can be obtained from (\ref{rkk2}). 
Eq.~(\ref{RS}) provides the essence of this method of determining $\gamma$.
Although the $\pm$ sign in this equation seems to indicate a possible 
discrete ambiguity in the solution for $\gamma$, we have shown in the above
numerical examples that the three observables $\rkk, \ckk$ and $\skk$ 
determine the weak phase in a unique manner. This follows from the 
requirement that $|T'/P'|$ lies in the range (\ref{T'/P'number}).

Since the direct asymmetry $\ckk$ occurs quadratically in Eq.~(\ref{RS}),
$\gamma$ is relatively insensitive to small values of this observable. 
Small values correspond to small values of $\delta'$ as predicted in 
\cite{BBNS}. The quantity $\rkk$, defined in (\ref{rkk}), is the
ratio of two charged averaged rates, the measurements of which do not require 
flavor tagging of neutral $\bs$ mesons. The error in this quantity is expected 
to be much smaller than in the mixing induced asymmetry $\skk$, which requires 
both flavor tagging and time-dependence. Therefore, for small values of 
$\rkk\ckk$, the dominant uncertainty in $\gamma$ originates in the error of 
$\skk$. 

As an illustrative example, suppose $\rkk = 0.85 \pm 0.05$, $\rkk \ckk \le
0.1$, and $\rkk \skk = 0.25 \pm 0.05$.  We find that
$\gamma$ is indeed insensitive to $\rkk \ckk$.  The
lowest value $\gamma = 50^\circ$ is attained when $\rkk = 0.8$ and $\rkk
\skk = 0.25$, while the highest value $\gamma = 70^\circ$ is attained when
$\rkk = 0.9$ and $\skk = 0.35$.  It can be checked that the values of $|T'/P'|$
for these solutions are within the allowed range.  Thus, an error on $\gamma$
of $\pm 10^\circ$, following from an error of $\pm 0.05$ in $\rkk\skk$,
requires a sample of several hundred $B_s \to K^+ K^-$ decays. 
A realistic number of required flavor tagged events,
including background and resolution factors,  is more like a few thousand.
Plans to measure $\bs(t) \to K^+K^-$ at the Tevatron Run II, with a comparable 
precision in $\skk$, are described in Ref.~\cite{Bworkshop}.

\section{Conclusions}

We have discussed time-dependent observables in the decay $B_s \to K^+
K^-$ in a manner parallel to that \cite{GR} for $B^0 \to \pi^+ \pi^-$.
Measurements based on the $B_s$ decay alone (when combined with a
measurement of the rate for $B_s \to K^0 \ok$) suffice to specify
both the weak phase $\gamma$ and the strong phase $\delta'$ which
is the difference between the penguin and tree strong phases.  The 
measurements are capable of resolving several discrete ambiguities.  In order
to achieve an error on $\gamma$ of $\pm 10^\circ$, it will be necessary to
obtain a sample of several thousand flavor tagged $B_s \to K^+ K^-$ decays. 

This method does not require full flavor SU(3) or discrete U-spin but only 
isospin symmetry.
We neglected two small contributions, a color-suppressed penguin amplitude in 
$\bs \to K^+K^-$, and rescattering effects in $\bs \to \ko \ok$, as well
as a small phase in $\bs-\obs$ mixing. Each of these terms is expected to
introduce an uncertainty in $\gamma$ of only a few percent. 

The observable $\dkk$, whose measurement depends on the observability of the 
width difference between the strange $B$ mass eigenstates, is predicted to
be large and positive.  (Its maximum value is $+1$.)
Any deviation from this prediction would be evidence
for physics beyond the standard Kobayashi-Maskawa picture of CP violation.

\section*{Acknowledgments}

This work was partially supported by the US - Israel Binational Science 
Foundation through Grant No. 98-00237.
The research of J. L. R. was supported in part by the United States Department 
of Energy through Grant No.\ DE FG02 90ER40560, while that of M. G. -- by
the Israel Science Foundation founded by the Israel Academy of Sciences and
Humanities. 

\def \ajp#1#2#3{Am.\ J. Phys.\ {\bf#1}, #2 (#3)}
\def \apny#1#2#3{Ann.\ Phys.\ (N.Y.) {\bf#1}, #2 (#3)}
\def \app#1#2#3{Acta Phys.\ Polonica {\bf#1}, #2 (#3)}
\def \arnps#1#2#3{Ann.\ Rev.\ Nucl.\ Part.\ Sci.\ {\bf#1}, #2 (#3)}
\def \art{and references therein}
\def \cmts#1#2#3{Comments on Nucl.\ Part.\ Phys.\ {\bf#1}, #2 (#3)}
\def \cn{Collaboration}
\def \cp89{{\it CP Violation,} edited by C. Jarlskog (World Scientific,
Singapore, 1989)}
\def \econf#1#2#3{Electronic Conference Proceedings {\bf#1}, #2 (#3)}
\def \efi{Enrico Fermi Institute Report No.}
\def \epjc#1#2#3{Eur.\ Phys.\ J.\ C {\bf#1}, #2 (#3)}
\def \f79{{\it Proceedings of the 1979 International Symposium on Lepton and
Photon Interactions at High Energies,} Fermilab, August 23-29, 1979, ed. by
T. B. W. Kirk and H. D. I. Abarbanel (Fermi National Accelerator Laboratory,
Batavia, IL, 1979}
\def \hb87{{\it Proceeding of the 1987 International Symposium on Lepton and
Photon Interactions at High Energies,} Hamburg, 1987, ed. by W. Bartel
and R. R\"uckl (Nucl.\ Phys.\ B, Proc.\ Suppl., vol. 3) (North-Holland,
Amsterdam, 1988)}
\def \ib{{\it ibid.}~}
\def \ibj#1#2#3{~{\bf#1}, #2 (#3)}
\def \ichep72{{\it Proceedings of the XVI International Conference on High
Energy Physics}, Chicago and Batavia, Illinois, Sept. 6 -- 13, 1972,
edited by J. D. Jackson, A. Roberts, and R. Donaldson (Fermilab, Batavia,
IL, 1972)}
\def \ijmpa#1#2#3{Int.\ J.\ Mod.\ Phys.\ A {\bf#1}, #2 (#3)}
\def \ite{{\it et al.}}
\def \jhep#1#2#3{JHEP {\bf#1}, #2 (#3)}
\def \jpb#1#2#3{J.\ Phys.\ B {\bf#1}, #2 (#3)}
\def \lg{{\it Proceedings of the XIXth International Symposium on
Lepton and Photon Interactions,} Stanford, California, August 9--14, 1999,
edited by J. Jaros and M. Peskin (World Scientific, Singapore, 2000)}
\def \lkl87{{\it Selected Topics in Electroweak Interactions} (Proceedings of
the Second Lake Louise Institute on New Frontiers in Particle Physics, 15 --
21 February, 1987), edited by J. M. Cameron \ite~(World Scientific, Singapore,
1987)}
\def \kdvs#1#2#3{{Kong.\ Danske Vid.\ Selsk., Matt-fys.\ Medd.} {\bf #1}, No.\
#2 (#3)}
\def \ky{{\it Proceedings of the International Symposium on Lepton and
Photon Interactions at High Energy,} Kyoto, Aug.~19-24, 1985, edited by M.
Konuma and K. Takahashi (Kyoto Univ., Kyoto, 1985)}
\def \mpla#1#2#3{Mod.\ Phys.\ Lett.\ A {\bf#1}, #2 (#3)}
\def \nat#1#2#3{Nature {\bf#1}, #2 (#3)}
\def \nc#1#2#3{Nuovo Cim.\ {\bf#1}, #2 (#3)}
\def \nima#1#2#3{Nucl.\ Instr.\ Meth.\ A {\bf#1}, #2 (#3)}
\def \np#1#2#3{Nucl.\ Phys.\ {\bf#1}, #2 (#3)}
\def \npps#1#2#3{Nucl.\ Phys.\ Proc.\ Suppl.\ {\bf#1}, #2 (#3)}
\def \os{XXX International Conference on High Energy Physics, Osaka, Japan,
July 27 -- August 2, 2000}
\def \PDG{Particle Data Group, D. E. Groom \ite, \epjc{15}{1}{2000}}
\def \pisma#1#2#3#4{Pis'ma Zh.\ Eksp.\ Teor.\ Fiz.\ {\bf#1}, #2 (#3) [JETP
Lett.\ {\bf#1}, #4 (#3)]}
\def \pl#1#2#3{Phys.\ Lett.\ {\bf#1}, #2 (#3)}
\def \pla#1#2#3{Phys.\ Lett.\ A {\bf#1}, #2 (#3)}
\def \plb#1#2#3{Phys.\ Lett.\ B {\bf#1}, #2 (#3)}
\def \pr#1#2#3{Phys.\ Rev.\ {\bf#1}, #2 (#3)}
\def \prc#1#2#3{Phys.\ Rev.\ C {\bf#1}, #2 (#3)}
\def \prd#1#2#3{Phys.\ Rev.\ D {\bf#1}, #2 (#3)}
\def \prl#1#2#3{Phys.\ Rev.\ Lett.\ {\bf#1}, #2 (#3)}
\def \prp#1#2#3{Phys.\ Rep.\ {\bf#1}, #2 (#3)}
\def \ptp#1#2#3{Prog.\ Theor.\ Phys.\ {\bf#1}, #2 (#3)}
\def \rmp#1#2#3{Rev.\ Mod.\ Phys.\ {\bf#1}, #2 (#3)}
\def \rp#1{~~~~~\ldots\ldots{\rm rp~}{#1}~~~~~}
\def \zpd#1#2#3{Zeit.\ Phys.\ D {\bf#1}, #2 (#3)}

\end{document}